# Superconductivity in the PbO-type Structure α–FeSe

Fong-Chi Hsu[1,2], Jiu-Yong Luo[2], Kuo-Wei Yeh[2], Ta-Kun Chen[2], Tzu-Wen Huang[2], Phillip M. Wu[3], Yong-Chi Lee[2], Yi-Lin Huang[2], Yan-Yi Chu[1,2], Der-Chung Yan[2], and Maw-Kuen Wu[2]

[1]*Department of Materials Science and Engineering, National Tsing Hua University, Hsinchu, Taiwan*

[2]*Institute of Physics, Academia Sinica, Nankang, Taipei, Taiwan*

[3]*Department of Physics, Duke University, Durham, North Carolina, USA*

**The recent discovery of superconductivity with relatively high transition temperature $T_c$ in the layered iron-based quaternary oxypnictides[1] La[ $O_{1-x}F_x$] FeAs was a real surprise. The excitement generated can be seen by the number of subsequent works[2-8] published within a very short period of time. Although there exists superconductivity in alloy[9] that contains Fe element, LaOM*Pn* (with M= Fe, Ni; and Pn=P and As) is the first system where Fe-element plays the key role to the occurrence of superconductivity. LaOM*Pn* has a layered crystal structure with an Fe-based plane. It is quite natural to ask whether there exists other Fe based planar compounds that exhibit superconductivity. Here we report the observation of superconductivity with zero resistance transition temperature at 8K in the PbO-type α–FeSe compound. Although FeSe has been studied quite extensively[10,11], a key observation is that the clean superconducting phase exists only in those samples prepared with intentional Se deficiency. What is truly striking, is that this compound has the same, perhaps simpler, planar crystal sublattice as the layered oxypnictides. Furthermore, FeSe is, compared with LaOFeAs, much easier to handle and fabricate. In view of the abundance of compounds with PbO type** **structure, this result opens a new route to the search for unconventional superconductors.**

FeSe comes in several phases: 1) a tetragonal phase α-FeSe with PbO-structure 2) a NiAs-type β phase with a wide range of homogeneity showing a transformation from hexagonal to monoclinic symmetry 3) an $FeSe_2$ phase which has the orthorhombic marcasite structure. The most studied of these compounds are the hexagonal $Fe_7Se_8$, which is a ferrimagnet with Curie temperature at ~125K, and monoclinic $Fe_3Se_4$.

Unlike the high temperature (high-Tc) superconductors[12] discovered more than 20 years ago, which has a $CuO_2$ plane that is essential for the observed superconductivity, the tetragonal phase α-FeSe with PbO-structure has an Fe based planar sublattice equivalent to the layered iron-based quaternary oxypnictides, which has a layered crystal structure belonging to the P4/nmm space group[1]. The crystal of *α*-FeSe is composed of a stack of edge-sharing $FeSe_4$-tetrahedra layer by layer, as shown schematically in Fig. 1. Polycrystalline samples with nominal concentration $FeSe_{1-x}$ (x=0.03 and 0.18) were synthesized and studied. X-ray diffraction analysis of the samples in Fig. 2 shows that α-FeSe is dominant and β-FeSe phases exist in trace amounts. The calculated lattice constants are a = 0.37693(1) nm and c = 0.54861(2) nm for $FeSe_{0.82}$, and a = 0.37676(2) nm and c = 0. 54847(1)nm for $FeSe_{0.88}$. The lattice constant slightly expands in the a-axis and shrinks in the c-axis for both samples as compared to those of *α*-FeSe in the JCPDS Card[13] (a = 0.3765 nm and c = 0.5518 nm). This lattice distortion should be most probably due to the deficiency of Se, and this gives a quick estimate of the difference in selenium concentration between $FeSe_{0.82}$ and $FeSe_{0.88}$.

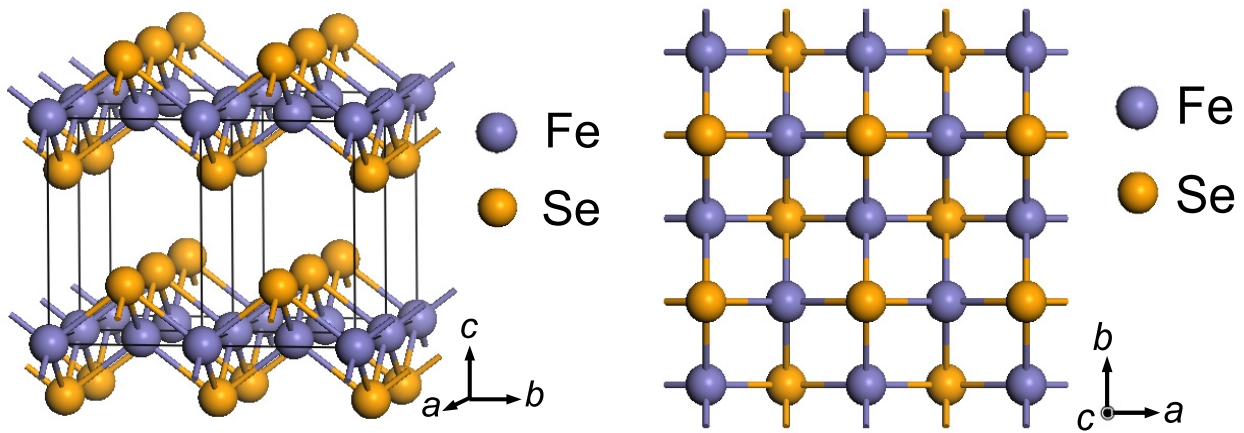

**Figure 1 | Schematic crystal structure of α-FeSe**. Four unit cells are shown to reveal the layered structure. Sample preparation was done in the following way: high purity (99%) powder of appropriate selenium and iron stoichiometry ($FeSe_{1-x}$ with $x \sim 0.03 - 0.18$) nominal compositions were mixed and grounded. The grounded powder was cold-pressed into discs with 400 kg/cm$^2$ uniaxial force, then the discs were sealed in an evacuated quartz tube and heat treated at 700 ℃ for 24hrs. The sample was reground and sintered again at 700 ℃ for 24hrs, then annealed at 400 ℃ for 36hrs. All the samples were kept in vacuum desiccators prior to measurement.

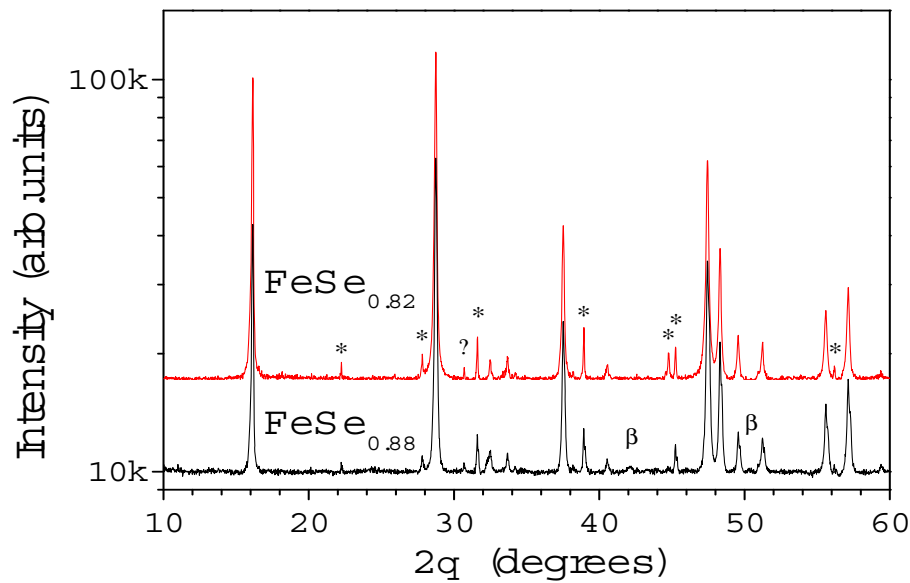

**Figure 2 | Powder X-ray Diffraction patterns of FeSe$_{0.82}$ and FeSe$_{0.88}$.** Phase identification of the sample was carried out with X-ray (Cu, K$\alpha$=1.5418Å) radiation in a Philip PW3040/60 diffractometer. The patterns show that the resulting sample with starting composition of Fe(53%)Se(47%) composes of primarily PbO-type tetragonal FeSe$_{1-x}$ (*P*4/*nmm*), the *α* phase, and partly of NiAs-type hexagonal FeSe (*P*6$_3$/*mmc*), the *β* phase. The sample with higher initial iron content, Fe(55%)Se(45%), shows no *β* phase but trace amounts of possible impurity phases including elemental selenium, iron oxide and iron silicide (marked as *). Question marks in the figure represent unknown phases. This is reasonable since in the Fe-Se binary alloy system, the *α* phase is considered as a slightly Se-deficient phase (45 to 49.4 at.% Se) and the *β* phase in contrast persists in a wide range of compositions from slightly Fe-rich to Se-rich (49.5 to 58 at.% Se)[13]. In FeSe$_{0.82}$, the possible iron oxide impurity phases could come from either starting materials (99.9% Fe) or surface oxidation during sintering, and the silicides might be the product of reactions between the sample and silica ampoules. Nevertheless, the samples contained only trace amounts of these impurity phases (note that the y-axis is in log scale).

Figure 3 displays the temperature dependence of electrical resistivity ($\rho$) of FeSe$_{0.88}$. The resistivity shows a broad bump at about 250K and exhibits metallic characteristics before the onset of the superconducting transition. The room temperature to residual resistance ratio is ~ 6. Interestingly, an anomalous downturn in resistivity at about 100K is observed. As the temperature is further lowered, the zero resistance transition occurs at about 8 K.

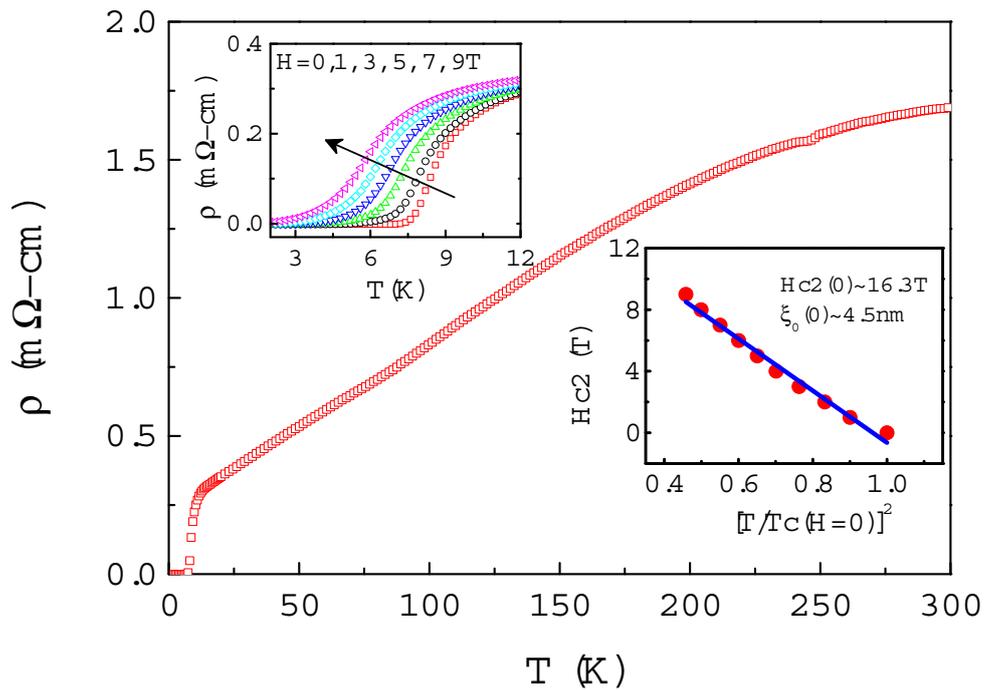

**Figure 3 | Temperature dependence of electrical resistivity (ρ) of FeSe$_{0.88}$.**
The sample resistance was measured using the standard 4-probe method using silver paste for contact. Measurements were carried out with a Quantum Design Physical Property Measurement System - model 6000 (PPMS). Superconducting critical transition temperature (Tc) in variety of applied magnetic field up to 9T was determined by taking the temperature where the resistance drops to 50% of that at the onset (50%Rn). The transition width is rather broad suggesting the inhomogeneous nature of the sample. The left inset shows the resistive measurement in magnetic fields (H) of 0T, 1T, 3T, 5T, 7T and 9T below 12 K. Tc decreases linearly with increasing magnetic field. The right inset displays the temperature dependence of upper critical field (H$_{c2}$), and the experimental result was fit to the relationship, $H_{C2}/H_{C2}(0) = 1 - (T/T_C(0))^2$, shown in blue line. The estimated upper critical field H$_{c2}$ (0) is about 16.3 Tesla; that gives a coherence length ξ$_0$ ~ 4.5 nm.

The magnetic susceptibility as a function of temperature was measured at 30 Gauss field strength, as shown in Fig. 4a. The zero field cool (ZFC) susceptibility is essentially temperature independent, indicating the sample is a Pauli paramagnet before the onset of superconductivity. A sharp drop indicating the magnetic onset of superconductivity appears at ~8K, which is the same as the zero resistance temperature. The susceptibility shows a relatively large Pauli susceptibility in the normal state. This relatively large positive background above Tc can possibly be attributed to the existence of trace of Fe impurity. Further confirmation of superconductivity is shown in the inset of Figure 4a, which displays the typical magnetic hysteresis curve for a superconductor. A relatively small magnetic anomaly occurs at about 105K, which is more pronounced in field-cool measurement. This magnetic anomaly occurs at the same temperature as that observed in the resistive measurement. Figure 4b shows the specific heat of the $FeSe_{0.88}$ sample. In the normal state, the derived electronic coefficient of specific heat $\gamma = 9.17$ mJ/mol-$K^2$. The superconducting transition at Tc is relatively broad and the estimated $\Delta Cp/\gamma T_c$ is ~ 0.61. In the superconducting state, the best fit of data over the temperature range 1 < Tc/T < 1.8 leads to $Ces/\gamma T_c = 7.69 \exp(-1.58\, Tc/T)$. But it does not follow the BCS relation over the full temperature range below Tc, as shown in the inset of figure 4b. This result may originate from the presence of impurity phases in the final product. However, it may also imply that the $FeSe_{1-x}$ is possibly an unconventional superconductor.

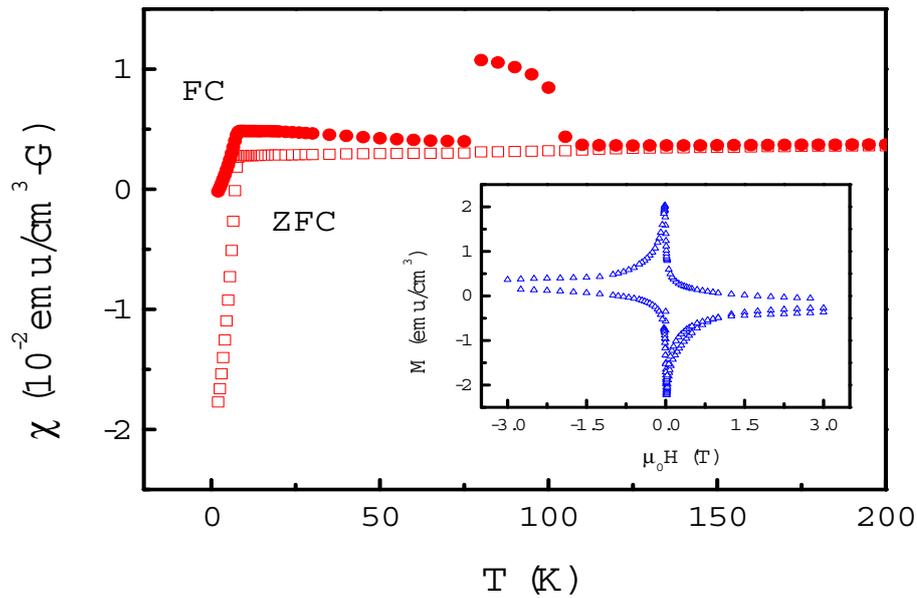

**Figure 4a | Temperature dependence of magnetic susceptibility measured in 30G magnetic field.** DC magnetic susceptibility measurements were performed in a Quantum Design superconducting quantum interference device vibrating sample magnetometer (SQUID-VSM). $FeSe_{0.88}$ powder sample is measured in two ways: 1) the sample is cooled without an initial external magnetic field applied (ZFC, open squares); and 2) then cooled in an initial external magnetic field (FC, solid squares). After initializing, a 30G magnetic field is applied and the susceptibility is measured as a function of temperature. Inset shows the magnetic hysteresis of the sample measured at 2K. It confirms the superconducting characteristic of the sample. A small magnetic anomaly is observed at about 105K, which is more pronounced in the FC measurements.

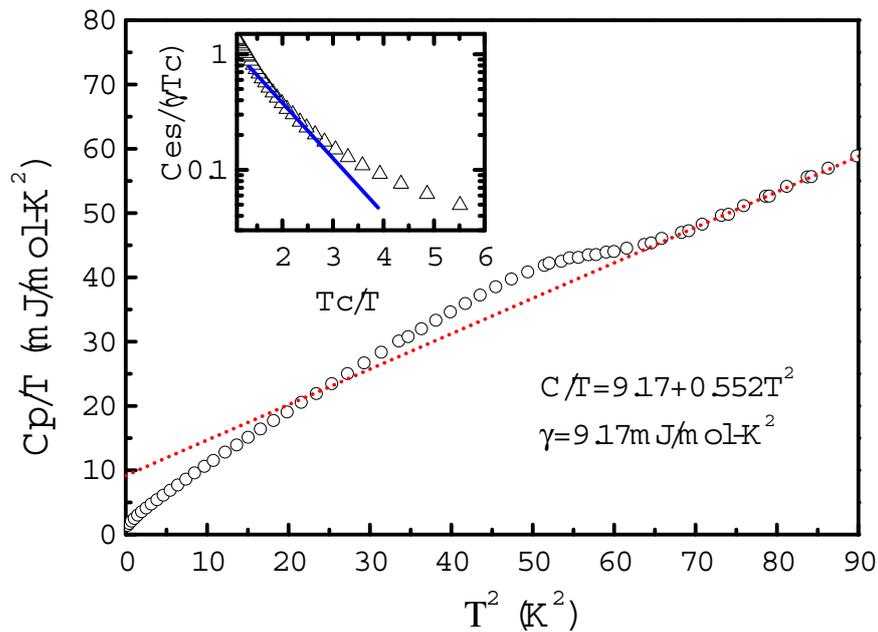

**Figure 4b | Low temperature specific heat of FeSe$_{0.88}$.** The measurement was carried out with thermal relaxation method in zero magnetic field. The red dotted line is the curve fitting of phonon and electronic contribution to the specific heat. The intercept at zero temperature gives γ = 9.17 mJ/mole-K$^2$. A specific jump appears at ~ 8K, which coincides with the zero resistance temperature that confirms the superconducting transition. The inset shows the semi-logarithmic Ces /γTc vs. Tc /T in the superconducting state. The measured plot displays deviation from linear curve of fully gapped superconductor (solid blue straight line).

Recently, it was reported[14,15] that there exists a pseudogap in La(O-F)FeAs, similar to that of the high Tc cuprates, based on the temperature dependent laser photoemission spectroscopy (PES). On the other hand, Mossbauer experiments[16] suggested that the suppression of magnetic and structural transition by F-doping in the LaOFeAs system may play the key role to the observation of superconductivity. It is possible that the anomaly observed in both the magnetic and resistive measurements in the FeSe$_{1-x}$ system may have similar origin. In fact, high resolution X-ray diffraction experiments at low temperature indeed show the presence of a structural transformation at ~ 105K. Figure 5 shows the diffraction patterns of the sample at different temperatures. Clear splitting of the diffraction peaks at ~105K is observed. Detailed refinement of the diffraction data suggests that the crystal structure changes from the tetragonal (P4/nmm) symmetry to the triclinic (P-1) symmetry. Whether this observed structural change at low temperature is related to the appearance of superconductivity is currently under intensive investigation.

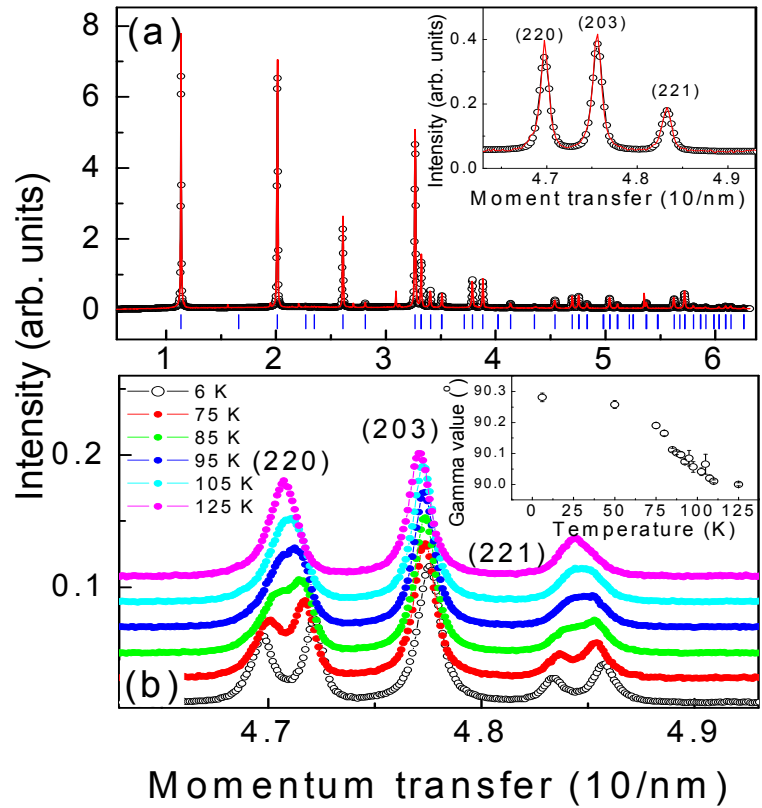

**Figure 5 | Temperature dependence of the high resolution X-ray powder diffractions for FeSe0.88 by synchrotron source at BL12b2 in SPring 8 with incident beam wavelength of 0.995 Å.** The moment transfer equals to M=4πsin(θ)/λ. (a) Observed (open black circle) and calculated (red solid line) powder diffraction intensities of FeSe0.88 at 300K using space group P4/nmm. The inset in (a) shows a single peak of the (2, 2, 0), (2, 0, 3), (2, 2, 1) reflection at room temperature. But double peaks show up for (2, 2, 0) and (2, 2, 1) at low temperature, as seen in (b). The double peak structure begins to show up at ~105K. The lattice parameters at 6K change to a = 0.3773 nm, b = 0.3777nm, c = 0.5503 nm; both angles α and β remain to be 90 degree but γ angle increases from 90 degree to ~ 90.3 degree. Inset in (b) shows the temperature dependence of $\gamma$ angle fit with P-1 symmetry.

**Acknowledgements** The authors like to acknowledge fruitful discussions with Prof. T.K. Lee and Prof. Sungkit Yip. We also thank the National Science Council of Taiwan and the US AFOSR through its Tokyo Office AOARD for their generous financial support.